ARTICLE

# Particle interaction with binary-fluid interfaces in the presence of wetting effects

F.K. Miranda* and C. Marchioli

University of Udine, Udine, Italy

*Corresponding author: marchioli@uniud.it

**Abstract**

In this paper we present an Eulerian-Lagrangian methodology for the simulation of the interaction between a fluid-fluid interface and a solid particle in the presence of wetting effects. The target physical problem is represented by ternary phase systems in which a solid phase and a drop phase interact inside an incompressible Newtonian carrier fluid. The methodology is based on an Eulerian-Lagrangian approach that allows for the numerical solution of the Continuity and Navier-Stokes equations by using a pseudo-spectral method for the carrier fluid, whereas the drop phase is modelled by the Phase Field Method (PFM), in which a smooth transition layer represented by an hyperbolic function is considered both across the solid-fluid interface and across the drop-fluid interface. Finally, the solid phase is described in the form of a virtual force using the Direct Forcing Immersed Boundary approach (DFIB). The properties of the immersed solid phase (including wetting effects), the deformability of the drops and the characteristics of the carrier fluid flow are the main controlling parameters that the method accounts for. To simulate a ternary phase system, the solid phase is coupled to the binary-fluid phase by introducing a single well potential in the free-energy density functional, which can also control the solid surface wetting property. The capabilities of the implemented tool are proven by examining first 2D and 3D validation case studies in which a solid particle is settling in a quiescent fluid. Then, the interaction of a solid particles with a binary-fluid interface and the effects of surface wetting on the submergence of a quasi-buoyant body are discussed. Finally, the equilibrium configuration for a solid particle interacting with an equally-sized drop at different contact angles and the relative rotation of two solid particles bridged by a drop are examined in the case the interaction is induced by shear fluid flow deformations on the drop interface.

## 1. Introduction

Particle and fluid interface interactions are ubiquitous in many natural and engineering systems, including emulsions, foams, and biological fluids. Understanding the behavior of these interactions is crucial for designing and optimizing various processes, such as microfluidics, drug delivery, wastewater treatment, filtering of gas exhaust pollutant and enhanced oil recovery [1, 2, 3, 4, 5, 6]. Numerical simulations have become an important tool for studying these interactions, allowing researchers to explore their behavior under different conditions and with different materials.

Particle behaviour in fluids can be simulated following a point-wise or size-resolved point of view. The point-wise particles approach represents particles are as mathematical points without any spatial extent. In this method, particles are treated as mass points with associated properties such as position, velocity, and mass. This simplified representation allows for efficient calculations and is commonly used when the size or shape of particles is not of primary importance for the simulation. Explicit interaction equations must be imposed to take into account the interaction between particles and binary fluid interfaces [7, 8].

Size-resolved particle simulations refer to a computational approach that explicitly takes into account the size,shape and and orientation of particles in the simulation. In size-resolved particle simulations, each particle is typically represented by a discrete volume or shape, such as spheres, ellipsoids, or irregular geometries. The size and shape of the particles are explicitly accounted for in the simulation, allowing for a more detailed characterization of their behavior and interactions with fluids and interfaces. However, this kind of simulations require careful consideration of computational resources, as the complexity and computational cost increase with the number and complexity of particles considered. Efficient algorithms and parallel computing techniques are often employed to tackle this challenge and enable large-scale simulations. Size-resolved particles simulations can be described using sharp or smoothed solid interface approaches.

Sharp interface approaches use Lagrangian point (tracers) to represent the topological shape of the particles and track their motion as they interact with the surrounding fluid, allowing for accurate modeling of complex fluid-particle and particle-particle interactions. For instance, the Discrete Element Method (DEM) models particles as discrete entities and considers their interactions based on contact mechanics principles. In this methods each particle is represented as a distinct entity with its own physical properties, such as size, shape, mass, and material characteristics. The motion and interactions of particles are determined by solving equations of motion for each individual particle. DEM enables the simulation of particle-particle and particle-wall interactions, as well as the study of particle segregation, mixing, and flow phenomena [9, 10, 11]. Another method widely used is the Immersed Boundary Method (IBM), it employs a force-coupling technique to represent the influence of particles on the fluid, while the fluid flow is solved on a fixed Eulerian grid. One of the key advantages of IBM is its versatility in handling different types of particles or immersed bodies, including rigid particles, deformable particles, or even biological cells. The method can accurately capture fluid-particle interactions, such as drag forces, lift forces, and boundary layer effects. It also allows for the investigation of complex phenomena, such as particle sedimentation, particle transport, or flow-induced deformations. However, IBM also has certain challenges. The force interpolation and back-coupling procedures require careful implementation to ensure accuracy and stability. The method can be computationally expensive, especially when simulating a large number of particles or complex particle shapes [12, 13, 14]. Finally, we can mention the Smoothed Particle Hydrodynamics (SPH), which is a meshless Lagrangian method that can be extended to simulate both fluid and particle phases. The fluid domain is discretized into a set of particles. Each particle carries information about fluid properties, such as density, pressure, and velocity. The simulation evolves by tracking the motion of these particles and updating their properties based on local interactions with neighboring particles. SPH naturally handles irregular particle shapes (rigid and deformable particles) and complex particle-fluid interactions [15, 16, 17].

Smoothed solid interface approaches use implicit advection equations to evolve the solid phase dynamics. They belong to the interface capturing type, i.e. a post-processing step must be done to retrieve the position and velocity of the solid interface at each time step. The solid interface is represented as a smooth transition layer. The solid body is represented then as a region with a distinct phase field parameter within the fluid domain. The phase field parameter is a continuous scalar field that smoothly transitions between values inside and outside the rigid body region. This allows for the description of the body's shape and motion without explicitly tracking its boundary. Usually the Phase Filed Method (PFM) is employed as a basis for smooth interface particles simulations, as we can see in the following references [18, 19, 20, 21, 22]. The performance of this approach is enhanced in particle-binary fluid interfaces interactions where the fluids interface is treated as a diffuse transition layer, especially when describing the dynamic contact line evolution, which does not require extensive and complex modeling, but most of the times is implicitly solved. Although this approach is relatively new and has usually a complex description, its applicability seems vast and promising, and the research interest in this technique is continuously increasing.

Recently, the development of new approaches on how to numerically simulate ternary phase systems involving binary fluids and solid surfaces interactions has received a great deal of attention. Based on their complexity, the approaches that are currently available in literature can be grouped into 3 main categories. The first category includes the approaches developed to treat flat wall boundaries. These are the ternary interactions simplest to implement, where the domain boundaries are treated as solid walls interacting with two distinct phases. The type of simulations allowed by this type of approach include, among others, channel flow laden with drops, bubbly flows in a tank and droplets impingement in flat surfaces [23, 24, 25, 26]. The second category includes the approaches developed to treat stationary arbitrary-shaped solid bodies. These approaches treat a static solid interface as a wall-boundary condition and are generally used to study problems such as porous media interactions with drops in a carrier fluid, drop impingement into curved surfaces, contact line evolution on a solid surface, meso-scale and macro-scale rigid structures immersed in a binary fluid [27, 28, 29, 30]. The third category includes the approaches developed to treat moving size-resolved solid particles in binary flows. In this case, the trajectory of the immersed particle can be altered by the fluid motion and by its own inertia. The studies carried out using this type of approaches are fully coupled and some of them present solid surface wetting effects. Therefore, their range of possible applications is vast, ranging from spheres sinking in water, buoyant bodies at water-air interface and sphere splashing into water to solid particles capture by drops and self-assembly induced by lateral capillary forces, just to name a few [31, 32, 18, 20].

Wetting effects play a crucial role in the behavior of particle-fluid interfaces. Wetting is defined as the ability of a liquid to spread or adhere to a solid surface, and is influenced by the solid surface characteristics. Understanding this effects is important for designing surfaces with desired wetting properties, such as superhydrophobic or superhydrophilic surfaces, for various applications. In recent years, several studies have focused on simulating size resolved particles-fluid interfaces interactions taking into account wetting effects. For example, Molecular Dynamic simulations have been used to investigate the behavior of droplets on superhydrophobic surfaces [33]. LB methods, perhaps the most popular approach for this kind of simulations, is used in several works [34, 35], to mention a few, in the simulation of thin film breakage on hydrophobic surfaces [36], lateral capillary forces on wettable cylindrical particles [37] and drop impact on cylinders at different contact angles. Phase Field methods have been also used to simulate wettable cylinders impacting on a free surface[18].

Although effective in simulating three phase interactions, the size-resolved particles methods usually need complex formulations and sophisticated numerical implementations. In this work, we present a simple and easy-to-implement numerical tool, where a Direct Numerical Simulation (DNS) of the incompressible carrier fluid flow is performed, the Phase Field Method describes the time evolution of the drop phase dynamics and the immersed solid particles are based on a hybrid Eulerian-Lagrangian description. These particles are tracked in a Lagrangian framework and their disturbance into the Eulerian domain of the fluid is spread using the Direct Forcing method. Their size and shape are bounded by a fictitious solid phase with a smooth interface. In addition, the wetting effects are also taken into account during ternary interactions. This allows to investigate two phenomena, neither of which have been previously numerically investigated to the best of the author's knowledge: the wetting effects in the submergence of a quasi-buoyant body and the relative rotation of two solids (bridged by a droplet), induced by shear fluid flow deformations on the drop interface.

## 2. Methodology

In this work the solid body trajectories are treated as point-wise particles in a Lagrangian framework. Each particle position is mapped in the Eulerian domain and linked to a region with a resolved shape and size of the corresponding solid body. Similar to You et al. [38] a Direct Forcing method is applied in this region, however, we describe the solid interface as a transition layer from the solid region to the fluid bulk using a smooth function in order to ensure the compatibility with the PFM [39, 19].

### 2.1   Single fluid and rigid-solid interaction approach

A generic incompressible Newtonian fluid flow is introduced as the carrier fluid flow, governed by the Navier-Stokes and continuity equations:

$$\nabla \cdot \mathbf{u} = 0, \tag{1}$$

$$\rho \left[ \frac{\partial \mathbf{u}}{\partial t} + (\mathbf{u} \cdot \nabla) \mathbf{u} \right] = -\nabla P + \mu \nabla^2 \mathbf{u} + \rho \mathbf{g} \tag{2}$$

Considering that the solid phase is described by a fictitious domain ($\psi_s$) built up by the union of $n$ individual body fields:

$$\psi_s = \bigcup_{i=1}^{n} \psi_i, \tag{3}$$

a phase parameter $\psi_s$ is inserted with constant values in the solid and fluid bulk volume ($\psi_s = 1$ and $\psi_s = 0$, respectively). The transition between phases is represented by a smooth layer, where fluid and solid properties coexist in proportions ruled by a hyperbolic tangential profile along the normal direction of the solid interface $\mathbf{x}$. In order to properly describe the local properties, the grid resolution must ensure the thinnest width with a well defined transition profile. Every individual rigid-solid sphere can then be generated using the following expression:

$$h(\mathbf{x}) = \frac{1}{2}[1 - tanh(\frac{\mathbf{x} - \mathbf{r}}{\xi_s})], \tag{4}$$

which is similar to the formulation used by Nakayama et al. [21], where $\mathbf{r}$ is the solid radius and $\xi_s$ is the parameter control for the interface width.

The fluid-solid coupling is achieved by adding a virtual force into the Navier-Stokes equations following the Direct Forcing Immersed Boundary approach [40, 41, 42]. In this method, the fluid within the solid region is enforced to follow prescribed solid-bodies velocities, ensuring the rigidity and the non-penetration condition [38].

The modified Navier-Stokes equations are:

$$\rho \left[ \frac{\partial \mathbf{u}}{\partial t} + (\mathbf{u} \cdot \nabla) \mathbf{u} \right] = -\nabla P + \mu \nabla^2 \mathbf{u} + \rho \mathbf{g} + \rho \boldsymbol{f}_{DF} \tag{5}$$

where virtf is the virtual force exerted in the solid region $\Omega_s$ to advance the solid object velocity from an intermediate time level velocity field $\mathbf{u}^*$ (where no influence of the solid is considered for its resolution) to $\mathbf{u}_s^{n+1}$ (calculated in previous steps) [38, 19, 43]. Eq. [6] shows how $\boldsymbol{f}_{DF}$ is calculated.

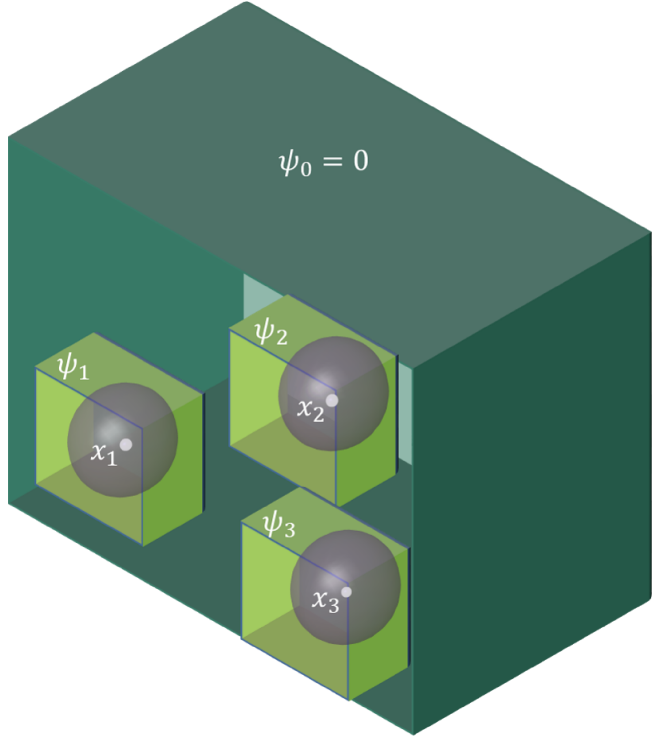

**Figure 1.** Solid phase generation diagram, satrting with the center body point mapping ($x_i$), subdomains calculation ($\psi_i$), continuing with individual body creation and finally merging sub-domains into a unique solid phase in the domain($\psi_s$).

$$f_{DF} = \frac{\mathbf{u}_s^{n+1} - \mathbf{u}^*}{\Delta t}, \tag{6}$$

where $\Delta t$ is the integration time step.

## 2.2 Equations of motion for a solid immersed in a fluid

The solid phase dynamics is described in a Lagrangian frame. The motion of an immersed rigid-body is caused by lineal and angular momentum. Consequently, the velocity $\mathbf{u}_s$ of the rigid-body can be decomposed in a translational and rotational velocity, as shown in eq. [7]:

$$\mathbf{u}_s = \mathbf{v}_s + \omega_s \times \mathbf{r} \tag{7}$$

where $\mathbf{v}_s$ is the immersed-solid linear velocity and $\omega_s$ its angular velocity with respect to the axis passing through its center of mass.
(*i*) We use the equation derived by Cheng-Shu et al. [38], shown in eq. [8], in order to obtain the lineal velocity of the solid object.

$$m_s \frac{d\mathbf{v}_s}{dt} = (m_s - m_f)g - \iiint_\Omega \psi_s \rho_f f dV + m_f \frac{d\mathbf{v}_s}{dt} \tag{8}$$

Where $m_s$ and $m_f$ are the solid and the fluid mass respectively, $g$ represents the gravity and $f$ is the average value of virtual force for the solid. The terms in the RHS of eq. [8] accounts on the effects of buoyancy, inertia and added mass (read from left to right). Subsequently this equation is discretized in time considering the following equivalences:

$$m_s = \iiint_{\Omega_s} \rho_s f dV = \iiint_{\Omega} \psi_s \rho_s f dV \qquad (9)$$

and

$$m_f = \iiint_{\Omega_s} \rho_f f dV = \iiint_{\Omega} \psi_s \rho_f f dV \qquad (10)$$

and for the virtual force term we use a 2$^{nd}$-order-accurate Adams-Bashforth scheme, obtaining as a result the following expression :

$$m_s \frac{\mathbf{v}_s^{n+1} - \mathbf{v}_s^n}{\Delta t} = (m_s - m_f)g - (\frac{3}{2} \iiint_{\Omega_s} \rho_f f^n dV - \frac{1}{2} \iiint_{\Omega_s} \rho_f f^{n-1} dV) + m_f \frac{\mathbf{v}_s^n - \mathbf{v}_s^{n-1}}{\Delta t} \qquad (11)$$

(*ii*) The angular momentum can be calculated from the intermediate time level velocity field $\mathbf{u}^*$ as follows:

$$\mathbf{J}_s \omega_s = \iiint_{\Omega_s} \rho_s \mathbf{r} \times \mathbf{u}^* \qquad (12)$$

where $\mathbf{J}_s$ is the rotational inertia of the solid body, $\mathbf{r} = \mathbf{x} - \mathbf{X}_s$ is the relative vector of a spatial point ($\mathbf{x}$) to the the center of mass of the solid body ($\mathbf{X}_s$). From eq. [12], we can calculate the angular rotation of the body center-of-mass.

Finally, the body trajectory is calculated by integrating the following expression:

$$\frac{d\mathbf{X}_s}{dt} = \mathbf{u}_s. \qquad (13)$$

### 2.3 Fictitious solid-phase with wettability in immiscible binary fluids

In order to include the wettability effects of a solids immersed in a binary fluid model, we modify the free energy density functional (eq.[??]), following the approach presented by Shinto [32]. This is based on the model of Cahn [44], who adds an additional surface term $\mathcal{F}_s$ (eq. 14) to describe the interactions between a binary fluid interface and a solid.

$$\mathcal{F}_s[\mathbf{X}_s, t] = \frac{1}{\beta} \int_S (-H\psi_s) dS \qquad (14)$$

where $\mathbf{X}_s$ is the position of the particle, $S$ is the particle surface and $H$ is a parameter which controls the wettability, we can precondition this property by tuning its value. For example in case of a fluid-drop system, where $\bar{\phi}_f = -1$ and $\bar{\phi}_d = +1$ represent the value of $\phi$ in the bulk of each phase, if $H = 0$, the solid surface is neutrally wettable, if $H < 0$, the solid has more affinity to the fluid, and if $H > 0$, the solid has more affinity to the drops. $\psi_s$ is the compositional order parameter of the solid. The binary fluid should evolve nearby this region, in order to accomplish the minimization of the free energy of the system [32, 28]. The tetaeq with respect to the affinity value can be calculated with the following expression:

$$cos(\theta_{eq}) = \frac{\mathcal{X}_S}{2}(3 - \mathcal{X}_S^2) \qquad (15)$$

with

$$\mathcal{X}_S = \frac{\bar{\psi}_s - \bar{\phi}_S}{\bar{\phi}_d - \bar{\phi}_S} \qquad (16)$$

and

$$\bar{\phi}_S = \frac{\bar{\phi}_f + \bar{\phi}_d}{2} \qquad (17)$$

where $\bar{\phi}_S$ and $\mathcal{X}_S$ describe the homogeneous solid surface and its affinity ($-1 \leq \mathcal{X}_S \leq 1$) [28]. A similar approach of implicitly imposing the contact angle by using an affinity parameter was developed by Guillaument et al. [24], who impose wetting effects using the penalty method.

The modified free energy functional considering the solid phase psis and the fluid phase $(1 - \psi_s)$ is shown in eq. [18]:

$$\mathcal{F}[\phi, \psi_s] = \frac{1}{\beta} \int_V d\mathbf{x}[f_b(\phi) + \frac{\kappa}{2}|\nabla\phi|^2 + \frac{K_s}{2}(\phi - \bar{\psi}_s)^2 \psi_s], \qquad (18)$$

where $\bar{\psi}_s$ is a constant value controlling the affinity, $K_s$ is a positive parameter (which has to be chosen as a large value compared with the parameters $\alpha$ and $\beta$) which ensures the value of the affinity inside the solid region in the phase field by imposing a single-well potential in the free energy functional [32].

The additional solid coupling term in the free energy functional makes the chemical potential (eq. [??]) evolve into the following expression:

$$\mu_\phi = \frac{\delta \mathcal{F}[\phi, \psi_s]}{\phi} = \alpha\phi^3 - \beta\phi - \kappa\nabla^2\phi + K_s(\phi - \bar{\psi}_s)\psi_s. \qquad (19)$$

In order to ensure the no-penetration condition, we employ the operator $(\mathbf{I} - \mathbf{n_s} \otimes \mathbf{n_s})$, which acts directly in the solid diffused interface, with $\mathbf{n_s} = \nabla\psi_s/|\nabla\psi_s|$ as the solid surface normal vector and $\mathbf{I}$ as the unit tensor. The advection-diffusion equation, taking into account the solid phase, results as follows:

$$\frac{\partial \phi}{\partial t} + \mathbf{u} \cdot \nabla\phi = \mathcal{M}_\phi \nabla \cdot [(\mathbf{I} - \mathbf{n_s} \otimes \mathbf{n_s})(\nabla\mu_\phi)]. \qquad (20)$$

## 2.4 Flow Field Equations

The equations that fully describe the incompressible flow of a generic Newtonian fluid with advected and deformable interfaces are the continuity equation (mass conservation) and the Navier-Stokes equation (momentum conservation) with an interfacial term (representing the coupling with the Cahn-Hilliard equation) and a virtual force term to account for the feedback of rigid-immersed bodies. The dimensional form of the mass conservation equation for incompressible flows is as follows:

$$\nabla \cdot \mathbf{u} = 0 \qquad (21)$$

In order to couple the two-phases-flow-field, we use a continuous approach to introduce boundary conditions at the interface [45, 46]. As for velocity, the transition at the interface should be continuous, avoiding sudden jumps, as shown in the following expression:

$$\mathbf{u}_1 \cdot \mathbf{n} - \mathbf{u}_2 \cdot \mathbf{n} = 0 \tag{22}$$

where $\mathbf{n}$ is the unit normal tensor to the interface and $\mathbf{u}_1$ and $\mathbf{u}_2$ represent the velocity vectors at each side of the interface. The jump condition for the stress tensor at the interface can be written as follows:

$$\mathbf{T}_1 \cdot \mathbf{n} - \mathbf{T}_2 \cdot \mathbf{n} = \mathcal{K}\sigma\mathbf{n} - \nabla_s \sigma \tag{23}$$

where $\mathcal{K}$ is the mean curvature, $\sigma$ is the surface tension, and $\mathbf{T}_1$ and $\mathbf{T}_2$ are the stress tensors at each side of the interface. The rhs of eq. [23] is composed by a normal ($\mathcal{K}\sigma\mathbf{n}$) and a tangential ($\nabla_s \sigma$) component, with $\nabla_s$ being the surface gradient operator.

The Navier-Stokes equations using the continuous approach in the binary fluid for a divergence-free velocity field is:

$$\rho(\phi)\left[\frac{\partial \mathbf{u}}{\partial t} + (\mathbf{u} \cdot \nabla)\mathbf{u}\right] = -\nabla P + \nabla \cdot \left[\eta(\phi)\left(\nabla \mathbf{u} + \nabla \mathbf{u}^T\right)\right] + \rho(\phi)\mathbf{g} \\ + \nabla \cdot [\bar{\tau}_c \mathcal{K}\sigma] + \rho(\phi)\mathbf{f}_{DF}(\psi_s), \tag{24}$$

with $\mathbf{u} = (u, v, w)$ as the velocity field, $\rho(\phi)$ and $\eta(\phi)$ as the local density and dynamic viscosity respectively, $\bar{\tau}_c$ as the Korteweg tensor, $\sigma$ as the surface tension and $\mathbf{f}_{DF}$ as the virtual force exerted by the solid phase.

## 2.5 Non-matched properties treatment

In order to avoid numerical discontinuities and jumps across the interface, the thermo-physical properties are defined to depend on the phase field indicator $\phi$ with smooth transitions across the interface.

We select arbitrarily the carrier phase ($\phi = -\sqrt{\beta/\alpha}$) as the reference property value, then the local density and viscosity are defined as:

$$\rho(\phi) = \rho_c \left[1 + \frac{\rho_r - 1}{2}\left(\frac{\phi}{\sqrt{\beta/\alpha}} + 1\right)\right] \tag{25}$$

$$\eta(\phi) = \eta_c \left[1 + \frac{\eta_r - 1}{2}\left(\frac{\phi}{\sqrt{\beta/\alpha}} + 1\right)\right] \tag{26}$$

with:

$$\rho_r = \frac{\rho_d}{\rho_c}, \quad \eta_r = \frac{\eta_d}{\eta_c} \tag{27}$$

where the subscript $d$ indicates the dispersed phase and $c$ the carrier phase.

We display two different dynamic viscosity ratios in fig. 2 ($\eta_r < 1$ and $\eta_r > 1$), which shows that the definition of the equations [25] and [26] prevent the value to reduce below zero (unphysical values).

## 3. Validation: Immersed solid interacting with a flat binary-fluid interface

In this chapter we study the interactions between a flat binary-fluid interface and a single solid. The first part (sect. 3.1) focuses on the evolution of the contact line along the 2D cylindrical surface at different wetting conditions. The following section (sec. 4.1) presents the simulation of a heavy cylinder sinking in a binary fluid system and the last part, sect. 4.2, shows the study of the wetting effects on the submergence of a quasi-buoyant cylinder in a binary fluid domain.

Here:

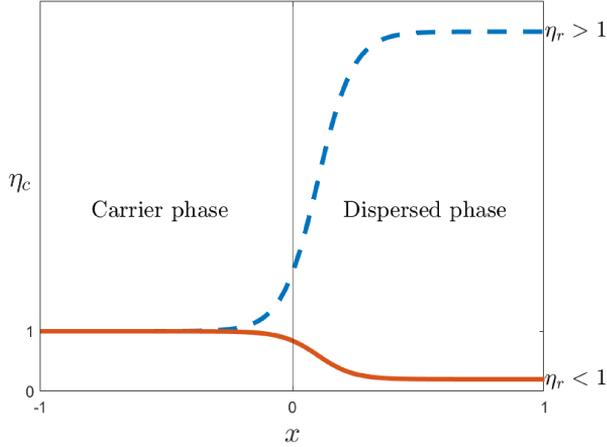

**Figure 2.** Transition profile of the dynamic viscosity when $\eta_r$ is greater than 1 (dashed-blue curve) and when $\eta_r$ is smaller than 1 (red-plain curve). The interface is identified by the vertical gray line.

### 3.1 Contact line equilibrium in curved surfaces

We perform the simulation of the contact line equilibrium for a cylinder in a binary fluid domain at different contact angles.

Similar to Shao et. al [27], our numerical setup consists of a squared domain of a binary fluid arranged as two horizontal layers with a cylinder fixed at the center of its interface (refer to fig. 3). The upper and lower boundaries are neutrally wettable walls and the left and right boundaries have a periodic boundary condition.

The cylinder has a radius of $r_{cyl} = 0.3h$ and the domain size is $2h \times 2h$ discretized with 128 grid cells along the periodic direction and with 257 grid cells in the wall normal direction. The lower fluid is set with the the phase parameter value of $\phi = 1$ and the upper one is set with $\phi = -1$. Depending on the wetting affinity preconditioned on the solid surface (eq. [15]), the contact line moves from the initial configuration (fig. 3) along the cylinder surface, until it reaches the equilibrium configuration. A hydrophilic surface with affinity $\chi = 0.35$, leads to an equilibrium contact angle $\theta_{eq} = 60°$ and the contact line reaches the equilibrium above its initial vertical position (see fig. 4). On the other hand, the contact line moves below the initial vertical position when the solid surface is preconditioned to be hydrophobic (affinity $\chi = -0.35$), leading to an equilibrium contact angle $\theta_{eq} = 120°$, while a neutrally wetting surface with affinity $\chi = 0$ leads to an equilibrium contact angle $\theta_{eq} = 90°$, where the interface remains flat and the contact line neither rises, nor goes below the initial position. This results match the theoretical angles predicted by eq. [15] and also the qualitative results obtained by other authors using different approaches [27, 36, 34].

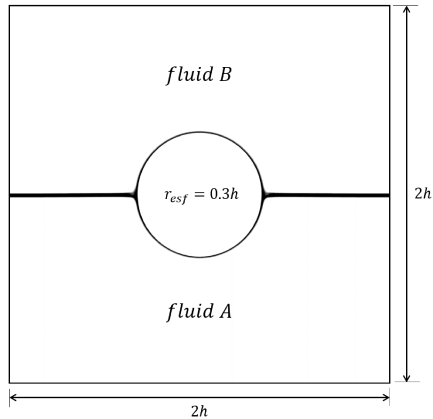

**Figure 3.** Initialization setup of the simulation domain.

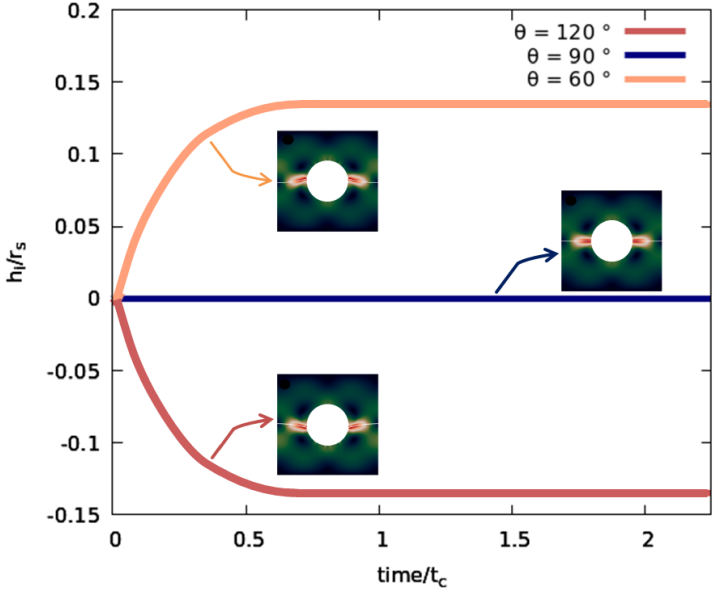

**Figure 4.** Equilibrium evolution of the contact line vertical position.

## 4. Results

### 4.1 Sinking of a heavy cylinder

Understanding the fluid dynamics developed by the motion of objects around a free surface is vital for some applications. In marine hydrodynamics, the wave loads produced by an immersed object motion, establish the basis of marine structures design, depending on the Froud number, the free surface motion may become violent and triggers several mechanisms like free surface breaking, cavity formation and cavity collapse [47]. A moving body in binary fluids is a complex scenario where the capability of numerical tools is tested in order to reproduce the fully coupled interplay of different parameters like surface tension forces, capillary forces, inertial forces and partial buoyancy forces (for a body migrating or trapped in between of two fluids).

In literature one can find numerical and theoretical studies of rigid-bodies motion near or through the free surface [48, 49, 50]. The one we are interested in this section is the simulation of a heavy cylinder sinking from the free surface of a binary fluid enclosed in a 2D tank. The parameters and conditions used here are taken from the experimental study carried by Vella et al. [51] .

Since the cylinder transits through two liquids, the partial buoyancy is taken into account, each time step, through the calculation of the solid portion immersed at each fluid, The numerical domain consists of a 2D tank filled with a fluid *A* (density $\rho_A$) up to $1.4h$ from the bottom wall, while the rest is filled with a fluid *B* (density $\rho_B$) as shown in fig. 5, where *h* represents half of the height of the domain and the fluids density and dynamic viscosity ratio are $\rho_r = \rho_A/\rho_B = 833.3$ and $\eta_r = \eta_A/\eta_B = 55.6$ respectively. As a first step, a cylinder with a radius of $r_s = 0.125h$ (neglecting gravity effects) is initially placed at the interface; using a contact angle of $\theta_{eq} = 105°$, the cylinder reaches its equilibrium position at $h_0 = 0.465$ (see fig.5 ). Then, turning on gravity effects and using a density ratio with respect to the fluid A of $\rho_s/\rho_A = 1.92$ the cylinder is released from $h_0$ respect to the free surface, letting the heavy cylinder sink. In order to compare with the experimental results [51], we use the characteristic time $t_c = (\sigma/\rho g^3)^{0.25}$, which is frequently used for gravity-capillary waves to travel the capillary length $l_c = (\sigma/\rho g)^{0.5}$ (the characteristic length), since the meniscus surrounding the cylinder is in hydrostatic equilibrium. We use a Reynolds number of $Re = 250$ and the Bond number is approximately equivalent to $Bo \approx (r_s/l_c)^2$. The latter value indicates that surface tension effects contribution is negligible for this experiment, and a neutral contact angle ($\theta_{eq} = 90°$) is thus assumed for the simulations.

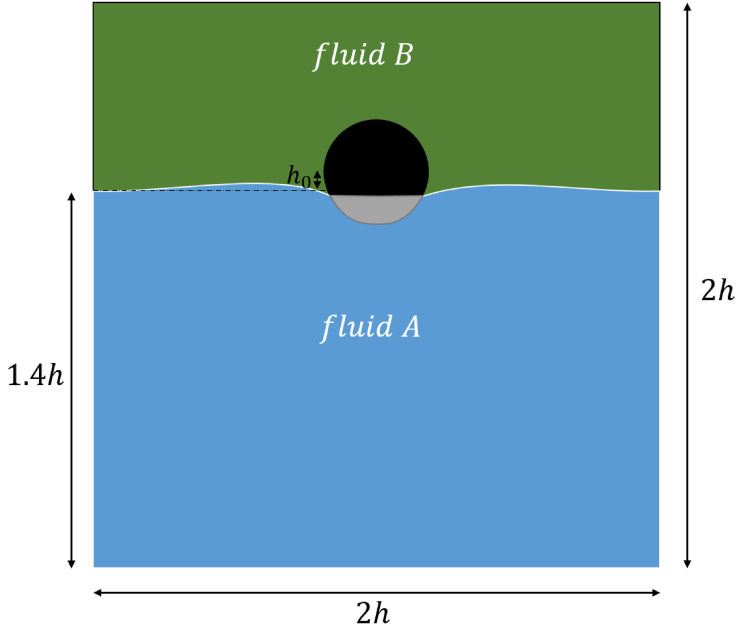

**Figure 5.** Scheme of the initial set-up for a cylinder with $\rho_s$ = 1920 $kg/m^3$ supported at an air-water free surface.

Fig. 6 shows a qualitative comparison of our simulations (right panel) against the experimental results (left panel), where we can observe that our simulation results are able to reproduce the main characteristic stages of the cylinder sinking experiment [51]: inflow in the region above the cylinder, cavity formation and jet generation [50].

The shape of the interface at the zone above the cylinder forms a kind of expander shape, which induces an upward jet of the fluid A and the entrainment of a portion of fluid B, attached to the cylinder surface, towards the bottom of the tank. This process is explained by fig. 7. Panel (a) shows that the cavity neck becomes narrow, squeezing the upper fluid out of it. The panel (b) and (c) then show how the neck walls merge, creating a pocket of upper fluid trapped and attached to the cylinder, which expands and coates the solid surface (neutral wetting), while the merged interface portion accelerates upwards, creating a bump above the free surface. When gravity and surface tension overcome the jet, the deformation is dissipated rapidly, turning the upward fluid motion into lateral waves —panel (d).

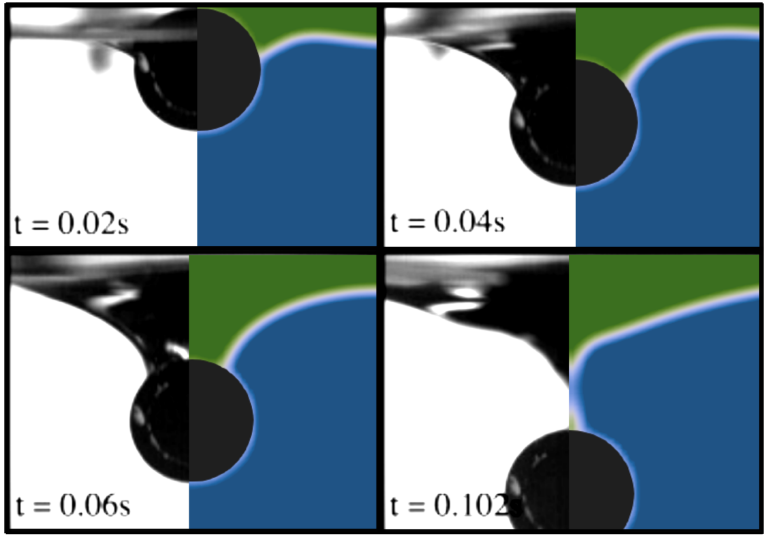

**Figure 6.** Time sequence comparison of the experiment (left panel) against our simulations (right panel) for a sinking cylinder of density $1920 kg/m^3$ in an binary phase system (green-blue region respectively). The cylinder is considered immersed when the cavity collapses.

For a quantitative comparison, we plot the results of the cylinder position evolution on time (both parameters non-dimensionalized with $t_c$ and $l_c$), as shown in fig. 8, where the numerical results, represented by a solid line, are plotted together with the experimental data [51] (represented by void circles) where a satisfactory agreement was reached.

The results obtained in this study case used a 2D domain with a periodic boundary condition in the horizontal direction and wall boundary condition in vertical direction. The grid independence test is performed to find the ideal mesh quality to efficiently reproduce experimental results. Three different mesh qualities are compared —coarse, medium and fine mesh (details can be found in table 1) using the processor model AMD Ryzen Threadripper Pro 3995WX @ $4.2GHz$.

Table 1 shows the result of the grid sensitivity test regarding the position evolution of the cylinder on time. We observe that the three curves overlap until time $t/t_c$ = 2.5, while the buoyancy and the added mass and the viscous effects are negligible. When the cylinder starts feeling the strong change in density and viscosity, the curves take different paths. The simulations result using a medium mesh quality are indistinguishable from the one obtained using a fine mesh, this indicates that the grid independence is reached using 128 × 256 grid cells. Therefore, the medium mesh quality is selected for further simulations with similar configurations.

**Table 1.** Parameters used in the simulations

| ID | Grid Resolution | Execution time per step [$s$] | Number of cores |
|---|---|---|---|
| Coarse | 64 × 128 | $35 \times 10^{-3}$ | 8 |
| Medium | 128 × 256 | $100 \times 10^{-3}$ | 8 |
| Fine | 256 × 512 | $280 \times 10^{-3}$ | 8 |

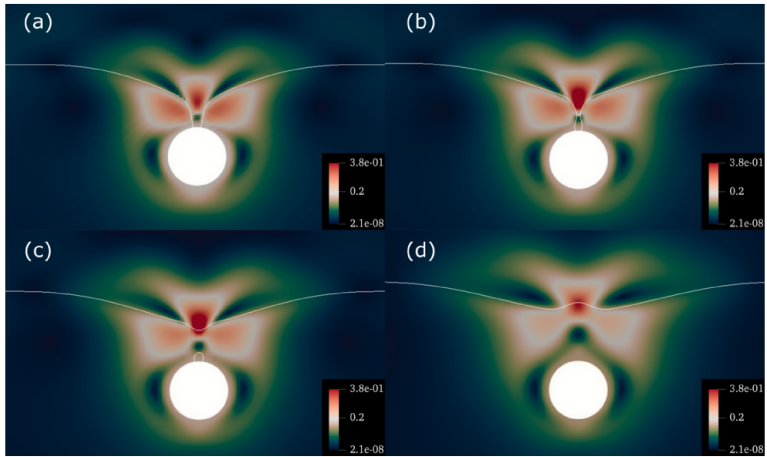

**Figure 7.** Sequence of jet formation after the cavity collapse, represented with the velocity magnitude field in the background. (a) upper fluid expelled from cavity; (b) cavity neck merged; (c) upper and lower jet formation; (d) jet rises the interface into the upper fluid.

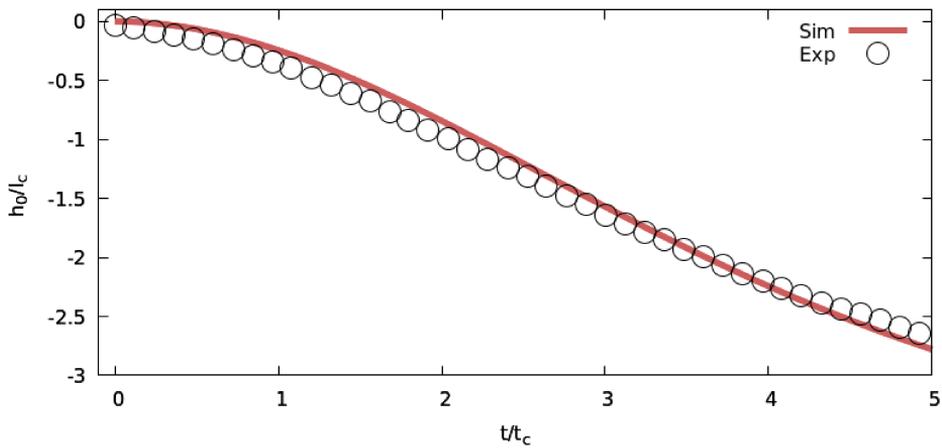

**Figure 8.** Simulation results of the sinking cylinder center position ($h_0/l_c$) evolution in time ($t/t_c$) compared with experimental data.

## 4.2   Submergence of a light cylinder in a binary fluid domain considering surface wetting effects

Let us consider the same geometrical configuration of the 2D domain employed in sec. 4.1, where the fluid A has a density of $\rho_{fA}$ = $1000 kg/m^3$. A cylinder (with slightly bigger density $\rho_s$ = $1130 kg/m^3$ than fluid *A*) released from the rest exactly at the interface (refer to configuration (a) in fig. 9b) would float indefinitely, because the sum of the capillary and buoyancy forces would exceed the cylinder weight [52]. In order to verify this statement, a simulation is carried out using the following parameters: fluids density ratio $\rho_r = \rho_A/\rho_B$ = 833.3, fluids viscosity ratio $\eta_r = \eta_A/\eta_B$ = 55.6, cylinder radius of $r_s$ = $0.125h$ and a neutral wettable solid surface.

The results are reported in fig. 10, where panel (a) shows the time sequence of the cylinder motion, and panel (b) shows its position oscillation over time, until it reaches the equilibrium and remains floating indefinitely.

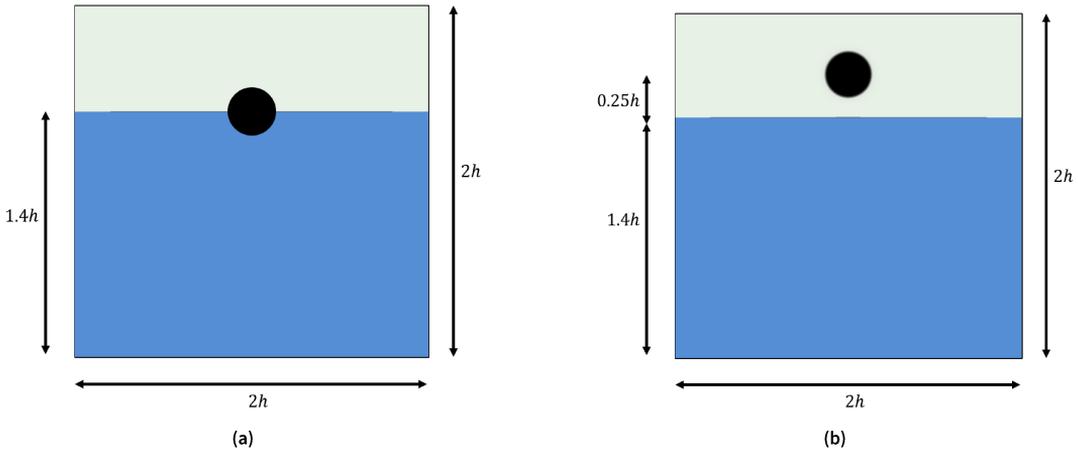

**Figure 9.** (a) Initial configuration for a floating cylinder. (b) Initial configuration for a cylinder submergence by adding inertia.

In order to submerge the cylinder, we increase its inertia by releasing it from a certain height above fluid interface as shown in fig. 9b, considering hydrophobic and hydrophilic solid surfaces. Once the solid is released, it starts accelerating because of the gravity, until it impacts with the free surface. Due to the strong contrast between fluid densities and the capillary component caused by the deflection of meniscus, the velocity then decreases.

Subsequently, the cylinder passes to the lower fluid, usually entraining a small drop (remaining from the fluid B after the breakthrough), which is stuck to the upper surface of the cylinder [52].

Figures 11 and 12 show qualitatively the difference in the immersion dynamics of two identical cylinders with different wetting conditions the first with contact angle of $\theta_{eq}$ = 70° and the second with $\theta eq$ = 110°. The first characteristic that attracts our attention is that the hydrophilic cylinder reaches a deeper position in the tank, generating an upward jetting stronger than in case of the hydrophobic cylinder. We can also observe how the velocity wake from the hydrophobic object is dissipated fast due to the resistance imposed by the surface tension. We may as well realize that the interface inflection due to the acute contact angle in fig. 11 helps with the formation of an upper fluid pocket trapped above the cylinder surface. Similar results for wetting objects sinking can be found in literature [18, 53, 54].

To further compare the wettability effects on the motion of rigid bodies entering a free surface, we plot the position evolution of the cylinder on time of both cases. Fig 13a shows that the hydrophilic cylinder reaches a deeper immersed position than the hydrophobic one. Fig. 13b shows that the hydrophobic solid decelerate from position $z_0/l_c$ 1.5 to 4.1 where finally the cylinder direction of motion is flipped. These results confirm that varying the wetting conditions in the cylinder surface, lead to important effects on the submergence dynamics and the three phase interactions. The wetting affinity of the solid surface towards one fluid, will thus control the ability of the body to submerge.

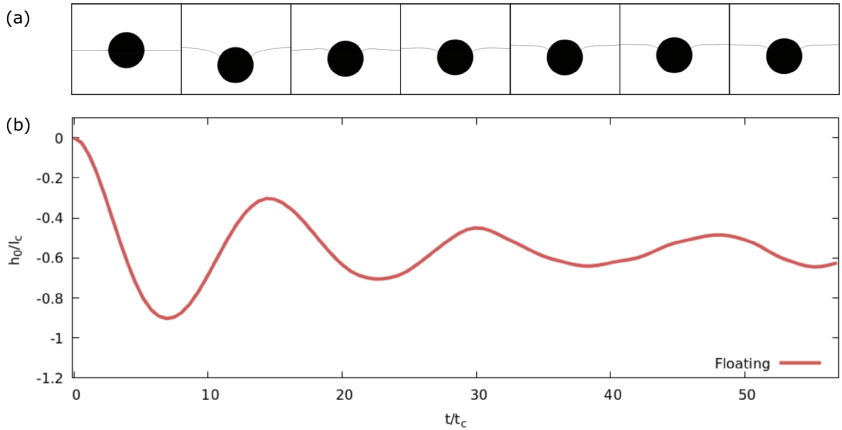

**Figure 10.** (a) Contact line evolution along the simulation. (b) Equilibrium evolution plot of the center position for a floating disk initialized at the free surface.

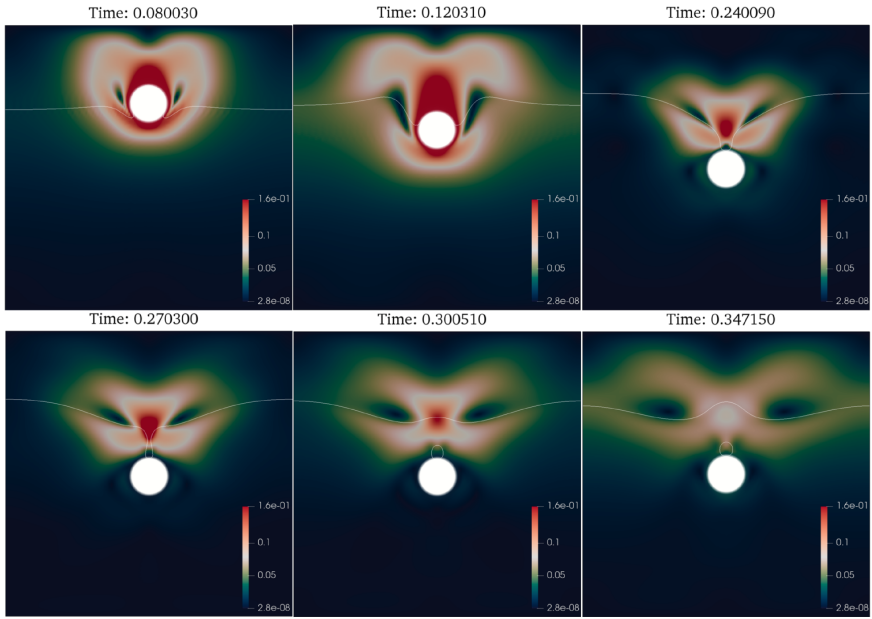

**Figure 11.** Time evolution of submerging hydrophilic disk interaction with the free surface depicted over the velocity field magnitude.

### 4.3  Wetting effects on the interaction of Solids and drops in three-phase systems

The type of binary fluids considered until chapter [3] were two stratified layers of fluids with an initial flat interface (free surface). In this chapter [4.3] we introduce a droplet as the second fluid phase immersed in a carrier fluid. In the first part, we study the evolution of the shape and position of a drop sitting on a cylindrical surface at different wetting conditions. The second part of the chapter considers two solid bodies interacting with a droplet of the same order of magnitude in a flow field. After the LBD has reached its equilibrium configuration , a shear field is initialized in the carrier flow and the LBD interactions are studied.

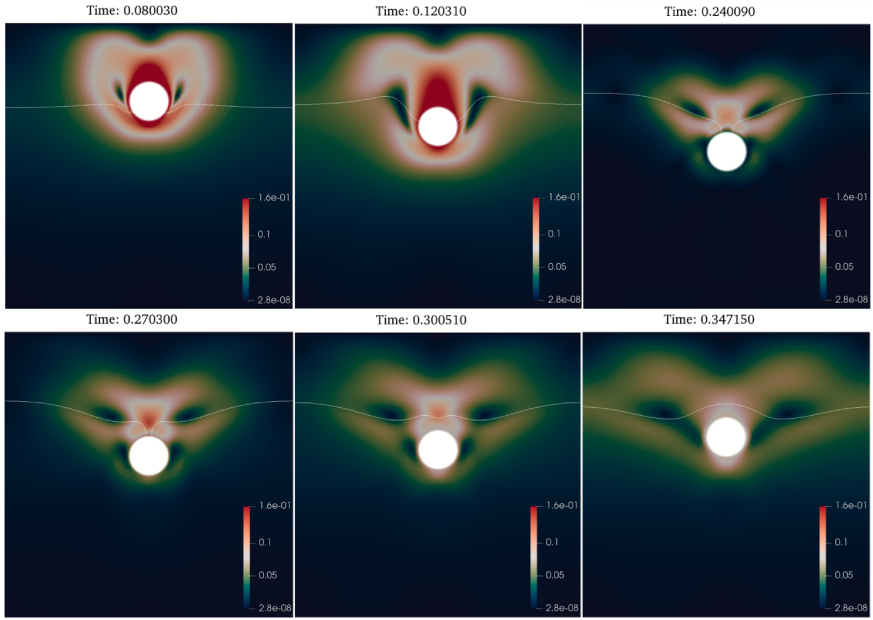

**Figure 12.** Time evolution of submerging hydrophobic disk interaction with the free surface depicted over the velocity field magnitude.

### 4.3.1  Solid-drop pair contact angle equilibrium

A complete work in solid-drop interactions is presented by Smith in [55], where several case studies are covered both experimentally (using one polystyrene sphere and one oil drop in an aqueous medium at different hydrophobic contact angles) and analytically (presenting equations for the prediction of the final equilibrium position for the pair in the range of 30° and 120° degrees of contact angle for various spheres-drop radius ratios). Another analytical formulation can be found in the wetting/dewetting section presented by Fakhari et al. [29], based on the premise of reaching the minimal free energy of the system by minimization of the peripheral area of a 2D drop sitting on a cylinder. These results obtained by means of the above mentioned formulation are used in the present work to validate the simulation of the dynamic contact line response to cylinder surfaces with changing wettability. Considering no gravity effects in the system during the simulations, we initialize our numerical domain by fixing a circular cylinder with radius $r_s = r_d = 0.33h$ at $0.67h$ from the lower wall. A drop with radius $r_d = 0.33h$ is placed at the center of the domain. We consider a fluid-drop density ratio of $\rho_d/\rho_f = 1000$, a viscosity ratio of $\eta_d/\eta_f = 100$ and a surface tension value of $\sigma = 0.01 N/m$.

The simulation is performed in a 2D domain with a wall boundary condition in the upper and lower limits and a periodic boundary condition in the side limits. A grid sensitivity study is carried out using the parameters listed in table 2 to determine the optimal mesh quality for the set of simulations with different contact angle values.

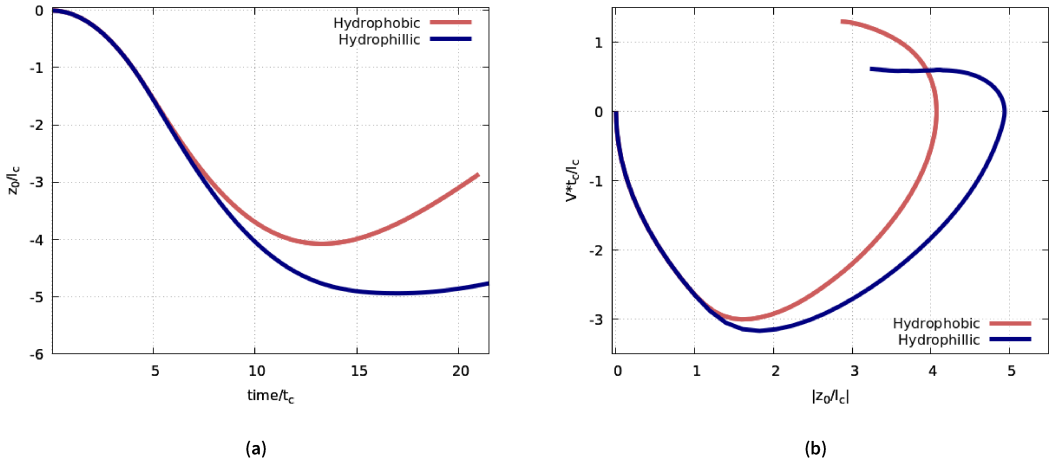

**Figure 13.** (a) Evolution of the cylinder velocity against the center position. (b) Evolution of the velocity against the cylinder position.

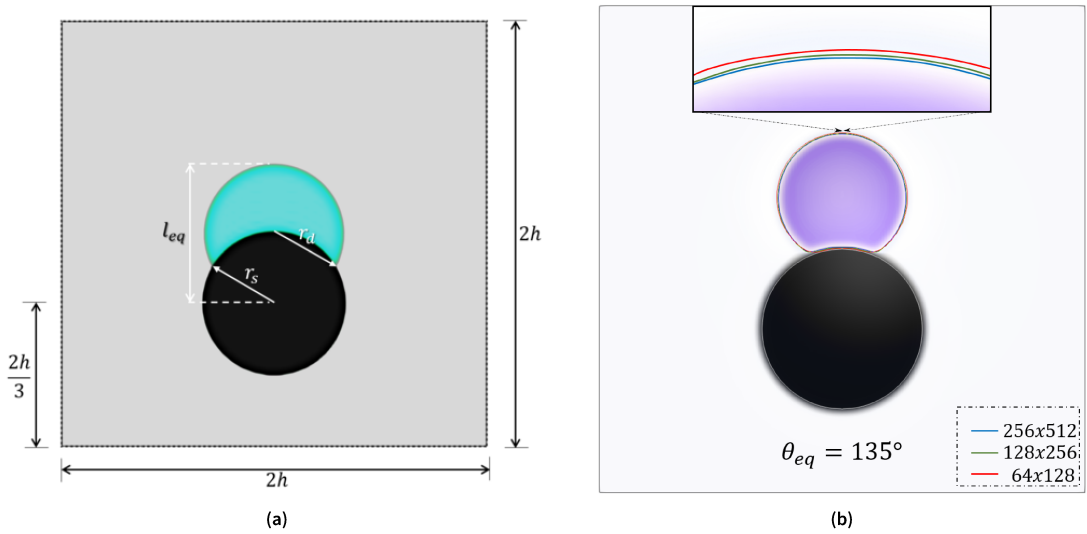

**Figure 14.** (a) Scheme of the initial configuration for the simulations. (b) Grid independence test for a drop-cylinder pair with a contact angle of $135°$ at equilibrium position, using: $64 \times 128$ (red line), $128 \times 256$ (green line) and $256 \times 512$ (blue line) grid cells in $y \times z$ direction respectively.

**Table 2.** Mesh quality list and numerical details for the simulation of the equilibrium configuration of a cylinder-drop pair.

| ID | Grid Resolution | Execution time per step $[s]$ | Number of cores |
|---|---|---|---|
| Coarse | $64 \times 128$ | $28 \times 10^{-3}$ | 8 |
| medium | $128 \times 256$ | $56 \times 10^{-3}$ | 8 |
| fine | $256 \times 512$ | $87 \times 10^{-3}$ | 8 |

Fig. 14b shows the drop interface (iso-surface at ϕ = 0) using different mesh qualities. One can observe as well that the grid independence is reached with 128 × 256 grid cells, where the difference between the drop interfaces using medium and fine mesh qualities is imperceptible.

Once the optimal mesh quality is found, a set of 7 simulations is proposed. Starting from the initial configuration presented in the initialization scheme of fig. 14a, the system is brought to its equilibrium configuration for a range of different contact angles (from 45° to 135° as presented in [29]).

The final equilibrium configuration will be represented by the variable $l_{eq}$, which is defined as a distance from the center of the cylinder to the highest point of the drop (refer to fig. 14a).

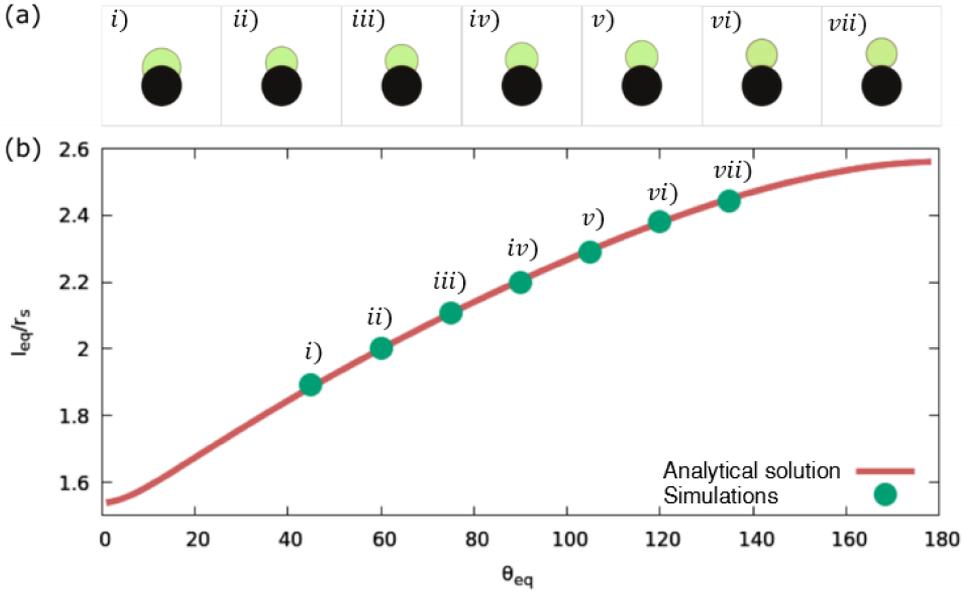

**Figure 15.** (a) Cylinder-drop pair equilibrium configuration at different $\theta_{eq}$. (b) Plot of the final configuration length $l_{eq}$ normalized by the cylinder radius $r_s$ for different contact angles and over the analytical curve result.

Fig 15a shows the results of the final equilibrium configuration of the cylinder-drop pair expressed in terms of $l_{eq}/r_s$ for each contact angle value used. Fig 15b shows the simulation results (represented by green markers) and the plot of the analytical solution (represented by the plain red curve). As illustrated in fig. 15b our results overlap almost perfectly over the analytical solution curve.

### 4.3.2   Liquid Bridged Doublets (LBD) in shear flow

In contrast to the above studied cases, this section includes a pair of free-moving solid bodies with active surface. They interact with a droplet in a carrier shear flow. The droplet deformation triggers the interplay between the inertial forces and the normal capillary forces, inducing the solids motion from an initial static state.

Several studies on the equilibrium configuration of bridged cylinders/spheres doublets can be found in literature [56, 57, 58]. They report analytical solutions in terms of the contact angle, the droplet volume and the maximal bridged doublet length $l_{max}$.
Further studies consider the interactions of a LBD in a shear flow field where the drop deformation and the contact line slippage may lead to a correlated rotation of the solids around a centered axis in the drop. Experiments on the mentioned relative rotation have been performed by Smith et. al

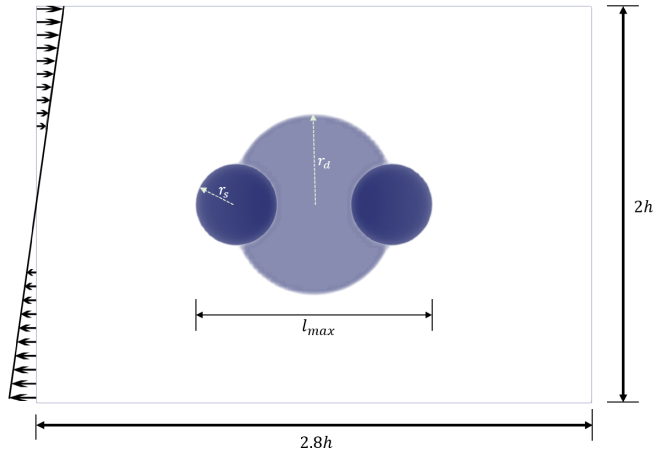

**Figure 16.** Schematic of the equilibrium configuration (represented by $l_{max}$) of a pair of hydrophilic disks bridged by a drop, and the shear flow field (represented by arrows) on which they are initialized.

[55], where three configurations of LBD (using spheres) were presented varying the relative volume of the bridging-drop with respect to the spheres.

In this thesis — due to constraints with computational resources and time, this study has been limited to a simplified 2D numerical experiment with matched density and matched viscosity for the fluids, but the LBD geometric proportions are based on an intermediate bridging-drop volume defined in [55]. The intention of this study is to ensure the solid surface wetting parameters, the bridging-drop properties and the shear flow field definition in order to induce a relative rotation [55].

Two identical hydrophilic ($\theta_{eq}$ = 73°) disks of radius $r_s$ = 0.20$h$ and density $\rho_s$ = 1120 $kg/m^3$ bridged by a droplet (immiscible liquid) of radius $r_d$ = 0.43$h$ and density of $\rho_d$ = 990 $kg/m^3$ were initialized and brought to equilibrium by the surface forces — neglecting gravity effects, obtaining a maximal distance of $l_{max}$ = 1.18$h$. The LBD system in equilibrium is then released in a shear flow field using a capillary number value of 0.24 (fig. 16). This simulation is carried out using 128 × 128 grid nodes in the directions $y \times z$ respectively, leading to a computational time step of $t_s$ = 32.8 × $10^{-3}$ $sec$ using 8 cores with the processor model AMD Ryzen Threadripper Pro 3995WX @ 4.2$GHz$.

Fig. 17 shows the phases of the rotation of the LBD in a shear flow field. The solid motion is started by the bridging-drop deformation into an elliptical shape as observed in panel (b) and (c), then, due to the shear field, the disks are accelerated horizontally in opposite directions, while the drop capillary forces keep them at the interface, giving them a vertical component of motion. This interplay results in a relative rotation of the disks and constant deformation in the topology of the bridging drop (since the solids are of comparable order of magnitude with the bridging-drop). Panels (d) and (e) show the asymmetry of the rotation respect to the vertical axis. While in the former, the bridging-drop seems to compressed, in the latter it seems to be elongated. The LBD reaches the maximal stretch when the disks reach the ellipsoidal vertices of the bridging-drop — panel (f). There, the solid inertial forces pull the bridging-drop in the directions of the disks against the normal capillary forces. As the pair continues rotating, the capillary forces overcome the inertial ones, bringing them closer and decreasing the deformation of the bridging-drop. The rotation is symmetric for the horizontal axis. The period of rotation of the LBD configuration is $TG$ = 0.28 sec.

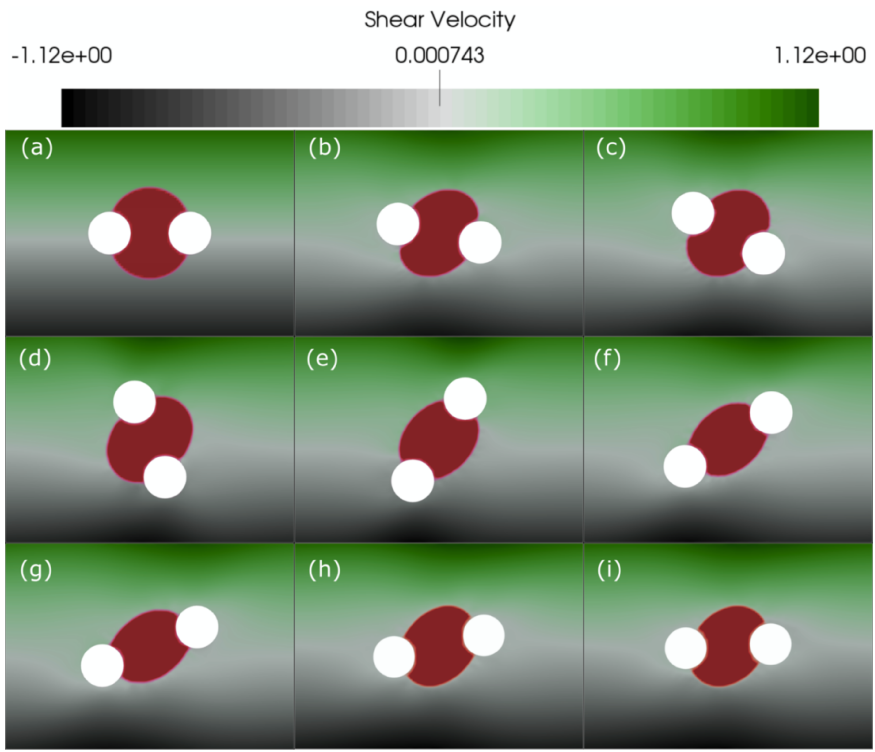

**Figure 17.** Sequence of the LBD rotating in a shear flow field. (a) Equilibrium position of the LBD. (b) The drop is deformed to an elliptical shape and the disks start moving. (c), (d) The elliptical shape is kept almost invariable, but the disks circulate along the drop interface. (e) The disks reach the ellipse vertices. (f), (g) The interplay between the disks inertia and the surface tension stretches and elongates the droplet. (h) The drop adopts a more rounded shape. (i) The LBD reaches almost a mirrored version of initialization configuration.

# 5. Conclusions

The dynamics of the interactions between solid and binary fluid interfaces in an incompressible Newtonian fluid have been characterized using multiphase numerical techniques: the Eulerian approach for the continuous liquid phase, the Phase Field Method to describe the drop phase topology evolution and the Direct Forcing approach for the motion of solids description. A fully coupled ternary phase numerical solver was achieved by adding into the carrier liquid a surface tension term (resulting from the dynamic effects of the drop phase) and a virtual force (which plugs the effects of the solid phase dynamics in the carrier fluid), and by using a single well potential to bound the solid region in the free-energy functional of the binary fluid system.

The settling of an immersed solid in a quiescent fluid was investigated at different fluid properties. Two-dimensional and three-dimensional simulations were performed and satisfactorily validated with analytical and experimental data. The contact line evolution was studied in cylindrical surfaces at different wettability conditions. The results showed that the fluids interface was perturbed in different ways: climbing up the solid surface for the hydrophilic case, retreating downwards for the hydrophobic case and staying still flat for the neutrally wettable case. In the second part, we performed a simulation of the interaction between a sinking cylinder and a binary fluid interface. The simulation results matched with great accuracy the experimental data, validating the phenomena both qualitatively and quantitatively. In the last part of the section, we investigated the wetting effects on the submergence of a quasi-buoyant cylinder in a binary fluid domain. From the simulations we observed that the *capillary flotation forces* either help or resist in the submergence. For hydrophobic conditions, the solid reached shallow depths, for hydrophilic conditions, on the other hand, it sunk deeper and easier. These results are in agreement with experimental and numerical findings in literature. The final part of this paper introduces second fluid phase as a a droplet and not as a stratified layer (as the precedent case study). First, we study the evolution of the shape and position of a drop sitting on a cylindrical surface at different wetting conditions. The resulting individual equilibrium configuration of the pair is represented by a solid-drop pair length. This length is then compared with available analytical and numerical data, to which our results match remarkably. The second part is devoted to the study of two solid bodies interacting with a droplet (both sized with the same order of magnitude) in a flow field. After the lbd has reached its equilibrium configuration (represented by a LBD length) in a stationary fluid, the shear flow field is initialized. The interactions within the LBD are originated by the interplay of capillary bridging forces and shear flow field effects. These interactions bring the LBD system into a relative rotation similar to the ones observed experimentally and, to the authors' knowledge, this phenomenon has not yet been addressed numerically in literature.

A limitation of this numerical implementation is that the solid sub-field must be regenerated at every time step, which increases the CPU calculation effort as we increase the number of solid particles used in the simulation; nevertheless, this limitation can be amended using optimization strategies. The simulations of three phase interactions work in three-dimensions (3D) as fine as in 2D; some cases were tested using a 3D setup; unfortunately, meaningful results required dedication of more time and computational resources; consequently, they are not shown in this work.

The current version of the code considers the effects of the solid spheres/cylinders rotation as additional values in the solid linear velocity; thus, actual solid body rotation is not performed; however, it can be included at expenses of added computational costs. The carried out work allows a number of potential further developments in terms of computational efficiency of the solver and of modeling capabilities of the solver. From the point of view of the computational efficiency, the solid phase solver is currently designed to handle computations of dozens and even hundreds of solid particles in an optimal way. The parallelization strategy consists in the equitable distribution of the total number of particles tasks to be computed, among all the cores allocated for the computation. The simulation of larger amount of particles (i.e. thousands or millions) may reach a bottleneck in terms of computational speed. Therefore, for the distribution and the calculation of all the particles tasks to be computed, an optimization study using GPU parallel processing is proposed instead.

From the point of view of modeling capabilities, further developments concerning non-spherical solids dynamics, big solids in drop-laden flows and lateral capillary forces in three-phase flows are suggested in the following lines. Although it is true that for a great amount of applications, the solid bodies can be modeled as cylinders or spheres in three-phase systems, there are some others (especially for microscopic, mesoscopic and macroscopic solids) where the shape of the solids plays an important role in the dynamics of the whole system. A solid shape in the latter cases can affect several parameters directly (to mention a few: the solid rotational inertia, the after-collision bounce direction, the partial buoyancy forces and the capillary forces). Therefore, a study of the effects of arbitrary-shaped solids on the interaction with binary fluid interfaces is encouraged.

Another topic to investigate further are the effects of considering big free moving particles in drop-laden flows. The use of small particles (point wise particles) for the stabilization of emulsions are broadly studied, especially in the cosmetic industry (due to the increasing demand of surfactant-free products). On the other hand, study results on the interaction of comparable size immersed solids and drops in drop-laden flows are still scarce. We therefore propose a study of the wettability effects on drops coalescence and breakage of big free moving particles in drop-laden flows. Taking a closer look at the solid dynamics around the interface, we observe that the capillary forces are the main mechanisms driving the three-phase interactions. These forces are responsible for the solids self-assembly in two-dimensional structures, on the free-surface of a binary fluid system. Two solid particles attract or repel each other when their interface perturbations overlap. Although there are several numerical studies in the field, just a few of them can handle the lateral capillary forces implicitly and without an extra model. The aim of a future study would be to carry out simulations of the effects of wetting using two identical buoyant solids attached to an interface. The simulations results must be compared with the experimental data on lateral capillary forces to determine the level of accuracy of the numerical tool and decide if a model is needed. The above mentioned capillary forces, for instance, represent in nature a mean of motility for some insects like the Pyrrhalta nymphaeae larvae. This creature has a wetting body circumscribed by a contact line (in the liquid free-surface). Therefore, in order to advance to the highest meniscus located on the edge of the liquid vessel, the insect arches its endings, perturbing the interface and forming a meniscus. These interactions generate capillary attraction forces between the insect and the edge of the vessel. Based on this real case phenomena, a study of the interaction of a simple flexible wettable membrane with the fluid-fluid interface is encouraged.

## Acknowledgement

This work has received funding from the European Union's Horizon 2020 research and innovation programme under Marie Skłodowska-Curie grant agreement no. 813948 (COMETE).

## Author contributions

- K. Miranda: Data curation, Visualization, Methodology, Software, Investigation, Writing- Original draft
- C. Marchioli: Conceptualization, Methodology, Investigation, Writing- Original draft